# The Correlations of Scene Complexity, Workload, Presence, and Cybersickness in a Task-Based VR Game

Mohammadamin Sanaei[1], Stephen B. Gilbert[1], Nikoo Javadpour[1], Hila Sabouni[1], Michael C. Dorneich[1], Jonathan W. Kelly [1]

[1] Iowa State University, Ames IA 50010, USA

**Abstract.** This investigation examined the relationships among scene complexity, workload, presence, and cybersickness in virtual reality (VR) environments. Numerous factors can influence the overall VR experience, and existing research on this matter is not yet conclusive, warranting further investigation. In this between-subjects experimental setup, 44 participants engaged in the Pendulum Chair game, with half exposed to a simple scene with lower optic flow and lower familiarity, and the remaining half to a complex scene characterized by higher optic flow and greater familiarity. The study measured the dependent variables workload, presence, and cybersickness and analyzed their correlations. Equivalence testing was also used to compare the simple and complex environments. Results revealed that despite the visible differences between the environments, within the 10% boundaries of the maximum possible value for workload and presence, and 13.6% of the maximum SSQ value, a statistically significant equivalence was observed between the simple and complex scenes. Additionally, a moderate, negative correlation emerged between workload and SSQ scores. The findings suggest two key points: (1) the nature of the task can mitigate the impact of scene complexity factors such as optic flow and familiarity, and (2) the correlation between workload and cybersickness may vary, showing either a positive or negative relationship.

## 1    Introduction

In recent years, the rapid progress in Virtual Reality (VR) technology has propelled its integration into diverse fields. In education and training, it offers immersive and safe learning environments for users to have effective learning outcomes [1]. Within the healthcare industry, VR provides potential benefits in medical practice, such as improving surgical skills, enhancing the patient experience, and facilitating efficient training for healthcare professionals [2]. Also, VR's immersive power in game industry can contribute to users' satisfaction in many cases such as presence, immersion, interactivity, and feedback [3].

This widespread adoption of VR applications in the mentioned domains underscores the growing need to gain a deeper understanding of the elements that shape user interaction and experiences within VR environments. Previous research has shown that VR environments may be different from video conferencing and face to face conditions in terms of social-related factors [4], task-related factors [5], and even performance [6]. These results necessitate new research to evaluate the relationships between different variables in the VR field to offer insights that could assist industry designers in creating better VR environments.

Several factors contribute to a person's overall VR experience: hardware, individual characteristics, and software [7]. Hardware factors refer to technology limitations of the device, such as screen resolution and field of view, along with customizable headset options, which can notably affect the experience [8]. Previous research also shows that individual characteristics such as age [9] and personality [10] can also affect VR experience.

Software factors, including scene complexity and the level of detail of the environments can also affect the engagement of the experience [11]. Virtual spaces with intricate and interactive content can provide a more realistic and compelling experience compared to simpler environments [12]. However, scene complexity can also impact the VR experience negatively; Fichna et al. [13] showed that high scene complexity can lead to increased cognitive load and visual clutter.

The significance of the relationship between scene complexity and VR experience is particularly pronounced in the domain of training, businesses, and virtual reality gaming, where users are not only entertained but are also presented with challenges that require the exertion of cognitive and motor skills. Thus, it is essential to study scene complexity in the game context and explore the relationship between scene complexity and other variables to help maintain user satisfaction in VR environments.

This research endeavored to explore the impact of scene complexity on participants' task-based VR gaming experiences, with a specific focus on four variables: scene complexity, workload, presence, and cybersickness.

## 2    Related Work

The relationship between scene complexity and workload could depends on multiple factors, such as the user's familiarity with the environment [14], the user's cognitive abilities [15], or time [16] High scene complexity can overwhelm users and cause cognitive load, making it difficult for them to understand and navigate through the environment [17, 18]. However, if users are adequately trained and adapted to the scene, they may become more efficient and experienced, leading to lower workload over time [14]. On the other hand, low scene complexity can result in a boring or less desirable experience [12]. So, the relationship between scene complexity and workload and their effects on VR experience can be non-trivial and dynamic. Therefore, in this study, the authors sought to contribute insight into how scene complexity and workload interact.

Another factor that might affect the users' VR experience could the relationship between scene complexity and presence [19, 20]. Although some studies have concluded that parameters of scene complexity like display resolution and texture mapping may not have a high influence on the sense of presence [21, 22], the findings of Witmer & Singer [23] and Welch et al. [20] on pictorial realism indicated that the higher level of details and realism in VEs led to an increased level of presence. Designing more complex scenes to have a better sense of presence may lead to higher possibility of cybersickness due to an associated increased perception of vection [19]. So, comprehending the dynamics of the relationship between scene complexity and presence is important for optimizing the VR gaming experience.

Understanding the relationship between workload and cybersickness has become another key research focus that could affect VR experience. Prior research has suggested that engaging in a task (higher workload) may contribute to reduced cybersickness in VR settings [24, 25]. Nevertheless, a recent study studying various workload levels and their impact on cybersickness indicated that participants in the higher workload condition had a significantly higher SSQ score and higher level of dropout possibility due to cybersickness [26]. Because the results seem conflicting, it's important to take another look at how workload and cybersickness are connected.

Presence is defined as observers' feeling of psychologically transitioning from the real world to the virtual world [27]. Understanding the relationship between presence and cybersickness has been one of the primary goals in achieving a successful virtual reality experience [28]. Some investigations have found a positive relationship between the presence level in VR and cybersickness [29, 30]. For instance, in a virtual grocery shopping task where participants were looking for a specific item, a strong correlation between presence and cybersickness were found [30]. Also, some studies have reported a null correlation [31, 32]. But the overall weight of evidence suggests that presence and cybersickness are negatively related [28]. As an examples, Sepich et al. (2022) showed that there is a significantly negative correlation between presence and cybersickness in one of the conditions of that study. Another study on an immersive multisensory environment, aiming to improve presence, found that subjective presence was negatively associated with participants' discomfort [33].

In addition, literature has sometimes shown a correlation between presence and task workload. A virtual pipe-cutting task designed by Draper & Blair [34] showed a significant positive correlation between presence and workload scores [34], which aligned with a Ma & Kaber [35] study, where higher presence led to higher workload levels. On the other hand, some research offered contrasting outcomes. For example, Riley & Kaber [36] showed a potential negative relationship between presence and mental workload. In this particular study, however, it might have been due to the high difficulty of the task, leading participants to report high frustration levels through NASA-TLX. The high workload led participants disengaged and some left the task unfinished [36]. Also, earlier research has suggested that there is minimal to no correlation between these two factors [26, 37].

## 3    Experiment Design

Based on the independent variable in this study, scene complexity, the Pendulum Chair had two different scenes, which were called simple scene and complex scene (Figure 1). These scenes are different in two key aspects: optic flow and scene familiarity. Based on the measurements of Lucas & Kanade [38] algorithm used across the first 10-second duration of 20 participants in the simple scene and complex scene, the complex scene had roughly 10 times the optic flow of the simple scene. Also, regarding the familiarity, in a separate study, 53 participants completed a pairwise comparison survey that posed the question "Which of these scenes is more familiar?" for all possible pairings of four images from the simple scene and four images from the complex scene. Then, using the Bradley-Terry [39] and an independent samples t-test among familiarity strength of those eight pictures, participants rated that the familiarity of the complex scene is statistically significantly higher than a simple scene, $t(6) = -7.785$, $p < .001$, $d = -5.50$.

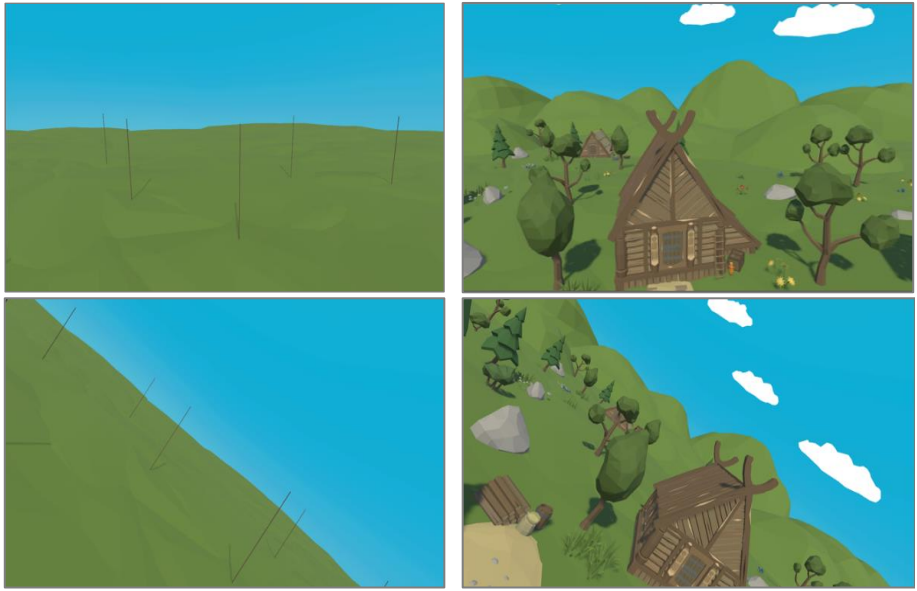

Figure 1. The simple and complex scenes.

Upon receiving approval from the Institutional Review Board (IRB), the researchers conducted their Pendulum Chair study data collection at Iowa State University. The study involved 44 participants consisted of 26 women, 16 men, one non-binary, and one preferred not to answer aged 18-53 ($M = 24.2$, $SD = 6.8$, $Mdn = 22.0$). All the participants had no history of photosensitive seizure disorders or vision problems, were fluent in English, and were at least 18 years of age. All the data was gathered from the first-person virtual reality (VR) game, Pendulum Chair, as described in Sanaei et al. (2023).

Half of the participants experienced the simple scene, while the other half experienced the complex scene. When participants arrived at the laboratory, they completed a pre-survey that included demographic information. Then, after having their interpupillary distance measured and the headset adjusted accordingly, each participants started to play the Pendulum Chair game [40]. In Pendulum Chair, the player sits in a chair atop an inverted pendulum and uses a joystick to prevent the chair from falling and hitting the ground (Figure 2). While the maximum of game play was 650 seconds, participants could exit the environment whenever they wanted. After playing the game, the researcher asked participants to fill in the post survey consisted of SSQ, NASA-TLX, and Presence questionnaires.

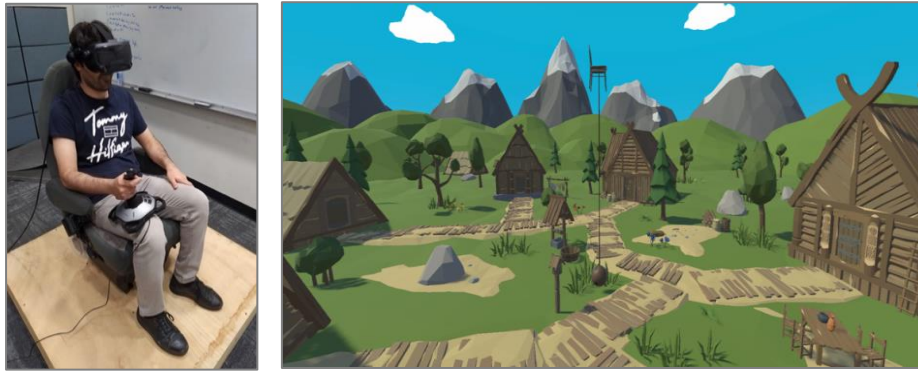

Figure 2. Participants' posture (left) and an external view of Pendulum Chair, showing the chair atop a pole that acts as an inverted pendulum. The goal is to stay balanced.

This study measured three variables as dependent variables in the Pendulum Chair. Post SSQ [41] was used to measure the cybersickness. The SSQ questionnaire contains 16 Likert scaled questions ranging from zero (no symptoms) to three (severe symptoms). To measure the workload that participants experienced, the authors used the NASA-TLX questionnaire [42] in the post survey. NASA-TLX consists of six questions which analyze participant's mental, physical, temporal demands, effort, performance, and frustrations in a 21-point scale for each of variables. Additionally, participants' feeling of presence in the Pendulum Chair was measured with an immersion questionnaire [43]. The immersion questionnaire consisted of 31 questions for which participants rated each using a 5-point Likert scale.

## 4      Hypotheses

The first hypothesis is written based on the Faure et al. [17] 's study, in which driving in more complex environments and performing a secondary task increased drivers' mental workload, as evidenced by higher blink rates and smaller pupil diameters. So,

H1 is that participants will experience higher workload in the complex environment. The second hypothesis is based on a study by Stanney et al. [19]. Stanney et al. [19] found that more complexity of the scene (in terms of more texture and details) enhanced the sense of presence for users, so H2 would be that participants experience a higher sense of presence in the complex scene. Next, per Sepich et al. [26]'s study where they found that more cognitive workload led to more sickness, the authors suggest in H3 that the higher workload participants experience, the more sickness they will have in the environment. Per Weech et al. [28]'s study, which conducted a comprehensive literature search about the relationship between presence and cybersickness and suggested a negative relationship between presence and cybersickness, H4 says that the higher sense of presence participants have, the less sickness they will experience in the environment. The last hypothesis is based on Draper & Blair [34] study, in which they found that higher presence led to higher workload levels; H5 predicts a positive relationship between workload and presence.

## 5    Results

Independent sample t-tests were used to analyze if there is any difference between the dependent variables based on condition (simple scene vs. complex scene). The assumptions of the t-test were analyzed by using inspection of a boxplot for assessing the outliers, Shapiro-Wilk's test for assessing the normality, and Levene's test for assessing the homogeneity of variances. Effect size for each comparison was measured using Cohen's $d$ to measure the effect size as large ($d = 0.8$), medium ($d = 0.5$), and small ($d = 0.2$) [44]. For correlations, a Pearson's correlation was used unless it was uncertain that the variable relationships were monotonic, in which case a Kendall's tau-b correlation was used.

If the two means in a comparison were very close, with a high p-value and a low effect size, an equivalence test was run to explore whether the two distributions were statistically equivalent. The two one-sided test (TOST) equivalence approach was used with independent sample t-tests [45] The null hypothesis was tested by setting the α level at .05. The key to equivalence testing is setting the upper and lower bounds for the equivalence range, or similarly, setting the smallest effect size of interest (SESOI). There are several different approaches to setting these bounds or SESOI. They include 1) basing them on a known standard or benchmark in the experimental domain, 2) basing them on an objectively measurable just-noticeable-difference, or 3) basing them on a subjective judgment of what difference would be of interest for a particular domain, as well as other approaches [46]. For example, in the domain of pharmaceutical drug equivalence, the U.S. Federal Drug Administration has designated that a generic drug must have no more than a 20% change from the original drug's mean efficacy to be deemed equivalent [47].

The present analysis used the last approach, as the authors conducted a thought experiment by asking themselves the question for each variable, "If someone said they had an intervention that would change [the variable] by X amount, how big does X need to be for the intervention to be of interest?" Multiple authors suggested that a 10% change from the mean would be worth noting and merit further exploration of this hypothetical intervention. Another viewpoint is that a change of 10% of the maximum value of the variable could be appropriate. Because of the subjectivity of this approach, strong claims about equivalence are not made, but equivalence testing was used as a tool for further exploring the relationship between the simple and complex scene data.

## 5.1 Workload

All t-test assumptions were met except that there was one outlier in the simple scene data. Analysis continued with the outlier because omitting the outlier did not change the result of the t-test (sensitivity analysis). There was no significant difference between mean of participants' workload experienced in the simple scene (10.89 ± 3.20) and mean of participants' workload experienced in the complex scene (10.55 ± 3.40), $t(42)$ = .342, $p$ = .734, $d$ = 0.103. Then, using bounds at 10% of the mean, ±1.07 did not yield a significant equivalence test ($p$ = .234). However, using bounds at 10% of the maximum amount of possible workload, ±2.1, an equivalence test showed that workload in both simple and complex scenes was statistically significantly equivalent ($p$ = .042) (see Figure 3).

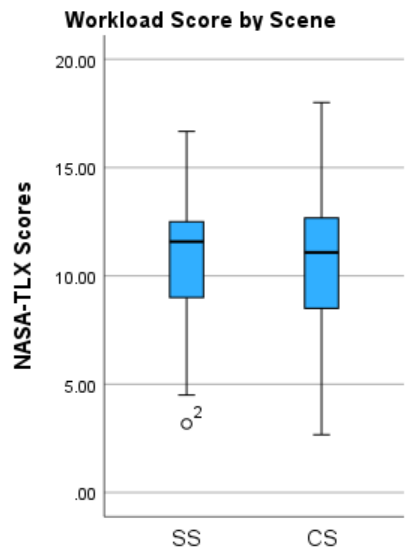

Figure 3. Boxplots of NASA-TLX workload scores for participants in the simple scene ($n$=22) vs. complex scene ($n$=22). Using 10% of the max NASA-TLX value for bounds yielded a significant equivalence.

### 5.2    Presence

All t-test assumptions were met. There was no significant difference between mean of participants' presence experience in the simple scene (98.54 ± 13.48) and mean of participants' presence experience in the complex scene (101.09 ± 19.17), $t(42) = .509$, $p = .613$, $d = 0.154$. Using bounds at 10% of the mean, ±9.98, yielded a near significant equivalence test ($p = .073$). However, using ±12.4 (10% of 155 − 31, the maximum and minimum presence score) yielded statistical equivalence ($p = .028$). The threshold 12.4 was 10% of the maximum amount of possible presence (Figure 4).

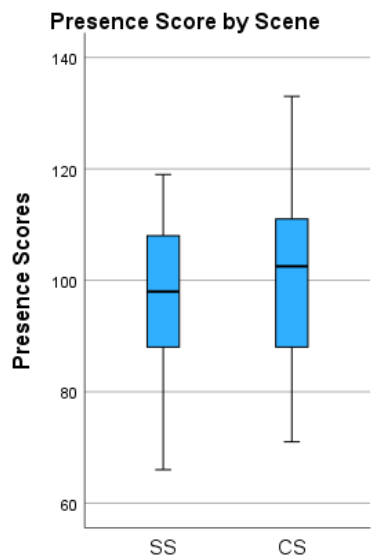

Figure 4. Boxplots of presence scores for participants in the simple scene ($n$=22) vs. complex scene ($n$=22). 10% of the max presence value for boundaries yielded a significant equivalence.

### 5.3    Cybersickness

The assumptions of the t-test were met. The result showed no significant difference between mean of SSQ score in simple (64.77 ± 40.53) and complex (52.02 ± 32.38) scenes, $t(42) = -1.153$, $p = .256$, $d = -0.35$. The closeness of the means of the Post-SSQ score across the simple and complex scenes suggested that it could be worthwhile to run a two one-sided test (TOST) equivalence test. Using an upper and lower bound of 10% of the mean (± 5.8), the distributions were not statistically equivalent ($p = .733$). However, the distributions were similar enough that using an upper and lower bound of ±32, 13.6% of the max SSQ value, yielded a significant equivalence effect ($p = .045$) (see Figure 5).

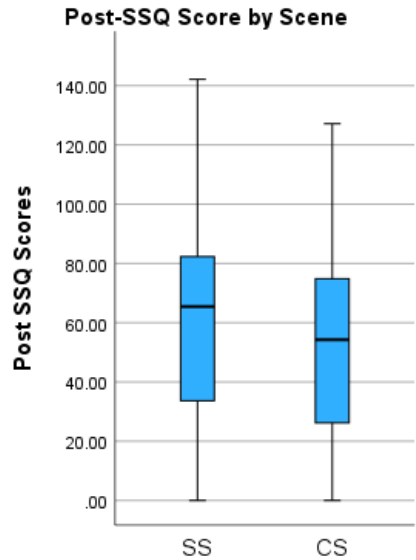

Figure 5. Boxplots of SSQ scores for participants in the simple scene (*n*=22) vs. complex scene (*n*=22). 13.6% of the max SSQ value for boundaries yielded a significant equivalence.

### 5.4    Correlations

There was a significant moderate, negative correlation between workload and SSQ scores, $\tau_b = -.38$, $p = .001$. The correlation between presence and SSQ scores was not significant, $\tau_b = -.87$, $p = .417$. The correlation between workload and presence was not significant, $\tau_b = -.05$, $p = .63$.

## 6    Discussion

H1 (higher workload in complex) and H2 (higher presence in complex) were not supported. Despite marked differences in optic flow and familiarity (scene complexity) between the simple and complex scenes, the observed equivalency, particularly within the 10% boundaries of the maximum possible workload and presence, prompts consideration of an additional influencing factor. The equivalent workload and presence in both scenes could be attributed to the uniformity of the task across environments. The presence of the same task in both the simple and complex scenes may result in comparable levels of workload and presence, potentially overshadowing the effects of optic flow and familiarity (scene complexity). This result suggests that the presence of a task, independent of scene content, has the potential to neutralize the impact of scene complexity factors.

H3 (higher workload correlated with greater sickness) and H4 (higher presence correlated with lower sickness) were not supported. Like optic flow and familiarity effects, similar outcomes may extend to the realm of cybersickness, where the observed equivalency hints at the potential dominance of factors beyond scene complexity, specifically workload and presence. Prior investigations have established correlations between cybersickness and workload [26] as well as cybersickness and presence [28]. The documented relationship between optic flow, a scene complexity factor, and cybersickness [7] indicates that increased optic flow often correlates with heightened cybersickness. Intriguingly, in the present study, despite significant differences in optic flow, no corresponding difference in cybersickness was observed. This could imply that the equivalency in workload and presence, stemming from the uniform task across scenes, might overshadow the impact of varying optic flow. It raises the possibility that the influence of presence and workload variables might outweigh the impact of optic flow in influencing cybersickness. A future focus on these variables could refine VR experiences, enhancing both user enjoyment and comfort.

Furthermore, the observed negative correlation between workload and cybersickness in the present study suggests that a higher workload imposed by the virtual environment may correlate with a reduced likelihood of experiencing cybersickness. While some studies have reported contradictory outcomes [26], Sepich et al. [26] demonstrated that increasing workload up to a certain limited threshold might result in decreased cybersickness—an alignment with our study's findings. The authors posit that the results of our study fall within this threshold, with the task potentially diverting individuals from cybersickness cues. These results underscore the need for future research to consider workload analysis, particularly exploring the maximum workload threshold that minimizes cybersickness.

The results did not reveal a significant correlation between presence and cybersickness. This lack of significance could potentially be attributed to the modest sample size in our study. It's plausible that with a larger sample size, the large negative correlation value could become statistically significant, aligning with existing literature that posits a negative relationship between presence and cybersickness. Moreover, the lack of support for H5 is consistent with findings from some previous studies that revealed minimal to no correlation between these two factors [26, 37]. This observation highlights the need for further investigation into this correlation in future research efforts.

## 7 Conclusion

In this study, the aim was to investigate the interplay between scene complexity, workload, presence, and cybersickness. Surprisingly, the results unveiled no significant distinctions between simple and complex scenes across any dependent variables. Equivalence testing, when considering 10% of the maximum variable value, further revealed statistical equivalence in workload and presence between both scenes. Cybersickness also demonstrated equivalence when employing 13.6% of the maximum SSQ value. As discussed in the discussion section, our findings suggest that the dynamics of scene complexity, workload, presence, and cybersickness may depend on the nature of the task users engage in within the VR environment. When users perform a task that is unrelated to the scene, scene complexity factors seem to be disregarded. Moreover, our study proposes that workload and presence might exert more influence than traditional factors such as optic flow in shaping cybersickness experiences. Despite significant differences in optic flow, our study found no substantial difference between simple and complex scenes when workload and presence were measured equal in both scenes. Intriguingly, these results challenge existing literature by indicating that an increase in workload correlates with a decrease in cybersickness, contrary to previous findings.

Future research should prioritize task considerations when examining cybersickness variables, as the nature of the tasks may potentially override the effects of other variables. Lastly, exploring various levels of workload in relation to cybersickness is a valuable avenue for further investigation. Such exploration can illuminate the threshold at which workload distracts users from cybersickness cues and may identify the workload conditions that may exacerbate the feeling of sickness in VR environments.